\documentclass[%
 aip,
 amsmath,amssymb,
 reprint,%
]{revtex4-1}

\usepackage{graphicx}
\usepackage{dcolumn}
\usepackage{bm}
\usepackage{epstopdf}
\usepackage{algorithmic}
\usepackage{amsmath,amssymb}
\usepackage{physics}
\usepackage{caption}
\usepackage{subcaption}
\usepackage{verbatim}
\usepackage{tikz}
\usetikzlibrary{arrows}
\usepackage{siunitx}
\usepackage[utf8]{inputenc}
\usepackage[T1]{fontenc}
\usepackage{mathptmx}
\usepackage{ulem}
\usepackage{standalone}
\usepackage[breaklinks=true]{hyperref}
\usepackage{breakcites}

\begin{document}

\preprint{AIP/123-QED}

\title[Rubrics for Charge Conserving Current Mapping in FEM-PIC]{Rubrics for Charge Conserving Current Mapping in Finite Element Particle in Cell Methods}

\author{Zane D. Crawford}
\altaffiliation[Also at ]{Department of Computational Science, Mathematics, and Engineering, Michigan State University, East Lansing, MI}
\author{ Scott O'Connor}
\altaffiliation[Also at ]{Department of Computational Science, Mathematics, and Engineering, Michigan State University, East Lansing, MI}
\author{John Luginsland}
\affiliation{Department of Electrical and Computer Engineering, Michigan State University, East Lansing, MI}
\author{ B. Shanker}
\affiliation{Department of Electrical and Computer Engineering, Michigan State University, East Lansing, MI}
\date{\today}

\begin{abstract}
Modeling of kinetic plasmas using electromagnetic particle in cell methods (EM-PIC) is a problem that is well worn, in that methods developed have been used extensively both understanding physics and exploiting them for device design. EM-PIC tools have largely relied on finite difference methods coupled with particle representations of the distribution function. Refinements to ensure consistency and charge conservation have largely been an \emph{ad-hoc} efforts specific to finite difference methods. Meanwhile, solution methods for field solver have grown by leaps and bounds with significant performance metrics compared to finite difference methods. Developing new EM-PIC computational schemes that leverage modern field solver technology means re-examining analysis framework necessary for self-consistent EM-PIC solution. In this paper, we prescribe general rubrics for charge conservation, demonstrate how these are satisfied in conventional finite difference PIC as well as finite element PIC, and prescribe a novel charge conserving finite element PIC. Our effort leverages proper mappings on to \emph{de-Rham} sequences and lays a groundwork for understanding conditions that must be satisfied for consistency. Several numerical results demonstrate the applicability of these rubrics.
\end{abstract}

\maketitle

\section{Introduction\label{sec:introduction}}

Modeling plasmas and other charged media as sources is a needed capability to better understand and design many real-world devices. Of the many techniques to model this phenomena, the particle-in-cell (PIC) method has been a popular approach. A wide variety of fields depend on PIC as a computational method, ranging from fundamental science of space weather, nuclear fusion, atomic processes on one hand, to high technology applications like particle accelerators, coherent electromagnetic radiation sources, and plasma processing devices on the other.  PIC, at a glance works as follows; charged media, represented as a collection of particles, move in a  domain under the influence of fields that are due to motion of these particles. This calls for self consistent solution of Maxwell's equations in conjunction with equation of motion. Given application drivers stated earlier, there has been extensive development of EM-PIC methods \cite{deca2014electromagnetic,lichters1997lpic++,gary2008cascade,chen2007electromagnetic}. Several of these have been used extensively in the design of devices \cite{mardahl2002intense,chakrabarti2003developing,fonseca2002osiris}. While traditional application were electrically large \emph{albeit} geometrically simple, there is burgeoning interest in devices with significantly geometric complexity but smaller footprint. As a result, there is need for EM-PIC methods with greater capabilities \cite{meierbachtol2015conformal}.

But before we delve extensively into nuances and needs of modern EM-PIC codes, it helps to understand a bit of history of challenges that have been overcome in developing EM-PIC methods. The principal challenge in EM-PIC modeling is self-consistency of solving Maxwell's equation together with equations of motion so as to actually capture dynamics of the distribution function. Challenges lies in incorporation of sources (and their variation over time) consistently into Maxwell's equation. As a result, there has been extensive work in understanding how one may develop methods that conserve charge \cite{chen2011energy,chen2014energy,sokolov2013alternating,glasser2020geometric}. Early development of EM-PIC methods largely relied on finite difference time domain (FDTD-PIC), as did addressing many of the aforementioned challenges \cite{vill,langdon}. More recent PIC development has seen extensive development mixed finite element methods (MFEM-PIC) and discontinuous-Galerkin (DG-PIC) as the EM field solver \cite{kraus2017gempic,na2016local,jacobs2006high}. 

These developments are specially pertinent as EM-PIC models that can more accurately capture geometry and physics with greater computational efficiency would soon be necessary to modeling physics ranging from novel accelerators to high frequency semiconductor devices. 
As the complexity of system to be modeled increases one needs EM-PIC suites that leverage advances in both finite element and particle methods. But as is to be expected, integrating the two calls for ensuring that self-consistency is preserved. This implies that discrete solutions of Maxwell's equations together with Newton's laws satisfy \emph{all} of Maxwell's equations together with the equation of continuity at every time step. This is easier said than done. Errors can manifest themselves as spurious charges and currents. To overcome these, it is possible to use procedures such as divergence cleaning to correct the erroneous fields that  arise. 
Approaches to correct the fields include using Lagrange multipliers \cite{marder}, solving Poisson's equation at multiple time steps \cite{boris,langdon}, or changing the field solver \cite{shadwick}. However, there can be a significant computational cost to using these methods. Of course, a more desirable alternative is to prescribe a rubric in which charge is conserved without correction terms or additional equations, regardless if sources are included or not. 

Developing computational methods that are inherently charge conserving has been a topic of interest as long ago as there have been EM-PIC codes \cite{buneman59,dawson62}. The benefits of having a charge conserving scheme are apparent; they avoid additional work necessary to effect divergence cleaning. Unfortunately,  their development has always been closely tied to a particular EM-solution technique, be in FDTD-PIC \cite{vill,east91} or MFEM-PIC \cite{pintoMap,moon,squire2012}. The latter is novel in that it develops a method that relies on \emph{de-Rham} maps to define the proper basis sets for representing the electric field and magnetic flux density (EB-MFEM-PIC). In effect, the primal quantities are defined based on Faraday's laws. The current mapping schemes demonstrate how they recover the results found in the early literature \cite{vill,east91}. As methods developed have been solver specific, significant attention has not been paid to examining the underlying mathematical structure and why that is necessary for charge conservation. In this work, we seek to do just that.

While unusual, we start our analysis at the particle approximation of the distribution function and develop a simple \emph{albeit} rigorous approach to demonstrate charge continuity relationships between charge density and currents. The exposition in this paper straightforward and uses the sifting property of the delta function. This is unlike that in presented in \cite{east91} that relies on an  discrete approximation. As an added benefit, this approach lays the groundwork for developing conditions for mapping particles and fields on any discrete grid, and consistently evolving their position and strength. In effect, we create a uniform framework and criterion that is necessary to solve all of Maxwell's equation consistently, thereby satisfying charge conservation as well. The rubrics prescribed are agnostic to the method used to solve Maxwell's equations. Indeed, we demonstrate how these are applicable to FDTD and (M)FEM. We also use a these rubrics to develop a novel MFEM-PIC scheme that uses electric flux density and the magnetic field (DH-MFEM-PIC) and validate its properties including charge conservation, satisfaction of Gauss' law as well as comparison against quasi-analytic data.

The rest of the paper is organized as follows: After defining the EM-PIC problem in Section \ref{sec:formulation} and its discetization using particle approaches, we define the three rules in Section \ref{sec:framework} in the context of preserving the spatial and temporal connections between Gauss' law and Ampere's law. 
In Section \ref{sec:spatialBasis}, we provide examples on how FDTD and two formulations of a variant of FEM, the mixed finite element method (MFEM), satisfy the three conditions. One of these is novel in that it relies on the electric flux density and the magnetic field (DH-MFEM). Next, in  Section \ref{sec:inconsistent}, we provide results that show what happens when the conditions are violated as well as affirm the correct behavior of MFEM when the conditions are met. In addition, several results are provided for the DH-MFEM-PIC that demonstrate necessary features for a number of examples (charge conservation, satisfaction of Gauss' laws, correctness against quasi-nalaytical data). In an Appendix, we present details of the MFEM discretizations and other proofs.

\section{Problem Statement and Formulation \label{sec:formulation} }

Consider a domain $\Omega$ whose boundaries are denoted by $\partial \Omega$. It is assumed that domain comprises of charged species that exist in a background medium defined by $\varepsilon_0$ and $\mu_0$, the permittivity and permeability of free space; for simplicity of the exposition, we consider only one species. It is also assumed that there exists an electromagnetic field, both impressed and arising from motion of the charged species. Both the fields and the charged species evolve in time. The evolution of the charged species is modeled as a conservative phase-space distribution function (PSDF) and those of fields using Maxwell's equations. To this end, we assume that the PSDF $f(t,\vb{r},\vb{v})$ must satisfy the Vlasov-Maxwell equation
\begin{equation}
\partial_t f(t,\vb{r},\vb{v})  + \vb{v} \cdot \nabla f(t,\vb{r},\vb{v}) + \frac{q}{m} [\vb{E} + \vb{v} \times \vb{B}] \cdot \nabla_v f(t,\vb{r},\vb{v}) = 0.
\end{equation}
The electric field $\vb{E}(t,\vb{r})$ and magnetic flux density $\vb{B}(t,\vb{r})$ satisfy Maxwell's equations
\begin{subequations}\label{eq:maxwell}
\begin{equation}\label{eq:far}
- \partial_t \vb{B}(t,\vb{r}) = \curl \vb{E}(t,\vb{r})
\end{equation}
\begin{equation}\label{eq:amp}
\partial_t \vb{D}(t,\vb{r}) = \curl \vb{H}(t,\vb{r}) - \vb{J}(t,\vb{r})
\end{equation}
\begin{equation}\label{eq:bgauss}
\div \vb{B}(t,\vb{r}) = 0
\end{equation}
\begin{equation}\label{eq:egauss}
\div \vb{D}(t,\vb{r}) = \rho (t,\vb{r}),
\end{equation}
\end{subequations}
 where fields and fluxes are related through constitutive parameters
\begin{subequations}
\begin{equation}
\vb{B}(t,\vb{r}) = \mu\vb{H}(t,\vb{r})
\end{equation}
\begin{equation}
\vb{D}(t,\vb{r}) = \varepsilon\vb{E}(t,\vb{r}).
\end{equation}
\end{subequations}
The moments of the PSDF yield 
the charge density $\rho(t,\vb{r})$ and current density $\vb{J}(t,\vb{r})$, defined as 
\begin{subequations} \label{eq:defChargeCur}
\begin{equation}
\rho(t,\vb{r})=q \int_\Omega f(t,\vb{r},\vb{v}) d\vb{v}
\end{equation}
\begin{equation}
\vb{J}(t,\vb{r})=q\int_\Omega \vb{v}(t) f(t,\vb{r},\vb{v}) d\vb{v}.
\end{equation}
\end{subequations}
 It follows from the above equations that charge and current densities are related via the equation of continuity

\begin{equation}\label{eq:curcont}
-\partial_t \rho(t,\vb{r})= \div \vb{J}(t,\vb{r}).
\end{equation}
Further, assume that these equations satisfy the initial conditions arising from 
\begin{subequations}
\begin{equation}
\div \vb{D} (0,\vb{r}) = \rho (0, \vb{r}) 
\end{equation}
and 
\begin{equation}
\div \vb{B} (0,\vb{r}) = 0
\end{equation}
\end{subequations}
To complete the description of the problem, we will assume that the
boundary of the domain $\partial \Omega$  may be partitioned into several subdomains, $\Gamma_i$, such that $\Gamma := \cup_i \Gamma_i$. Each $\Gamma_i$ has a boundary condition, either Dirichlet ($\Gamma_D$), Neumann ($\Gamma_N$), or impedance ($\Gamma_I$). The boundary conditions are defined as
\begin{subequations}\label{eq:bceq}
\begin{equation}
\hat{n}\times \mathbf{E}(t,\mathbf{r}) = \bm{\Psi}_D(t,\mathbf{r})\;\;\text{on}\;\Gamma_D
\end{equation}
\begin{equation}
\hat{n}\times \frac{\mathbf{B}(t,\mathbf{r})}{\mu} = \bm{\Psi}_N(t,\mathbf{r})\;\;\text{on}\;\Gamma_N
\end{equation}
\begin{equation}
\hat{n}\times \frac{\mathbf{B}(t,\mathbf{r})}{\mu} - Y\hat{n}\times\hat{n}\times \mathbf{E}(t,\mathbf{r}) = \bm{\Psi}_I(t,\mathbf{r})\;\;\text{on}\;\Gamma_I
\end{equation}
\end{subequations}
where $\hat{n}$ the vector normal to the surface $\partial \Omega$, $Y$ is the surface admittance defined as $\sqrt{\varepsilon/\mu}$, and $\Psi_D(\mathbf{r},t), \Psi_N(\mathbf{r},t), \text{and } \Psi_I(\mathbf{r},t)$ represent some function.
These equations describe a self consistent set that must be solved numerically. In what follows, we describe the numerics necessary to do so.

\section{Self consistent modeling framework
\label{sec:framework}}

While we do not solve the Vlasov system directly, we follow the usual practice of representing the PSDF using a collection of basis sets and then evolving the trajectory of these in using Newton's law in concert with Maxwell equations.  The basis sets are defined on a cell or simplical complex which discretizes a volume $\Omega$ bounded by $\partial\Omega$. The mesh consists of $N_g$ nodes, $N_e$ edges, $N_f$ faces, and $N_v$ volumes (in three dimensions).

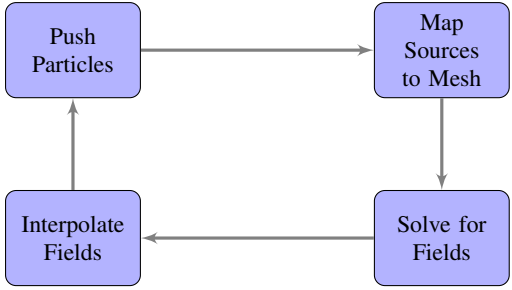
\begin{figure}
\centering
\tikzstyle{decision} = [diamond, draw, fill=blue!20,
    text width=4.5em, text badly centered, node distance=2.5cm, inner sep=0pt]
\tikzstyle{block} = [rectangle, draw, fill=blue!30,
    text width=5em, text centered, rounded corners, minimum height=4em]
\tikzstyle{line} = [draw, very thick, color=black!50, -latex']
\begin{tikzpicture}[node distance = 2.5cm, auto]
\node [block] (push) {Push Particles};
\node [block] (map) [ right= 4cm ] {Map Sources to Mesh};
\node [block, below of=map] (solve) {Solve for Fields};
\node [block, below of=push] (interp) {Interpolate Fields};

\path [line] (push) -- (map);
\path [line] (map) -- (solve);
\path [line] (solve) -- (interp);
\path [line] (interp) -- (push);
\end{tikzpicture}
\caption{Flow of typical EM-PIC simulation.}
\label{fig:PICcycle}
\end{figure}

This process of modeling the charged media can be described as the EM-PIC cycle as seen Fig. \ref{fig:PICcycle}. It begins with mapping the charged particles on the grid, forming a discretized representation of $\rho(t,\vb{r})$ and $\vb{J}(t,\vb{r})$. These discretized currents and charges act as sources in discretized Maxwell solvers, and are used to evolve the fields for one time step. The ``new'' fields and flux densities are then used to update the locations of the charges and thereby obtain currents (and charges). The cycle then repeats. 
What is apparent from the the above are the the nuances of what ensures self consistent modeling of Maxwell and Vlasov systems discrete grids; this is the focus of this Section. 

\subsection{Sampling the distribution function, and equation of continuity}

To begin, we start with representing the PSDF using
\begin{equation}
\label{eq:samplingPhase}
f \left (t, \vb{r},\vb{v} \right) = \sum_{p=1}^{N_P} S\left (\vb{r}-\vb{r}_p(t) \right ) \delta \left ( \vb{v} - \vb{v}_p (t)  \right )
\end{equation}
where shape functions, $S(\vb{r})$, represent the spatial support of particles. It follows from the definitions in \eqref{eq:defChargeCur} that $\rho(t,\vb{r})$ and $\vb{J}(t,\vb{r})$ can be represented as 
\begin{subequations}
\label{eq:sources}
\begin{equation}
\rho(t,\vb{r}) = q\sum_{p=1}^{N_p}S(\vb{r}-\vb{r}_p(t))
\end{equation}            
\begin{equation}\label{eq:Jsource}
\vb{J}(t,\vb{r}) = q\sum_{p=1}^{N_p}\vb{v}_p(t)S(\vb{r}-\vb{r}_p(t)).
\end{equation}
\end{subequations}
Given the discrete representation of the PSDF, it can be shown that the equation of continuity still holds in this discrete/particle setting. While this has been proven by Eastwood \cite{east91}, the approach we take is straightforward, and exploits the sifting property of the delta function. Specifically, we use the definition $\vb{r}_p(t) = \vb{r}_{0p}(t) + \int_0^t d\tau \vb{v}_p(\tau)$ and the identity
\begin{equation}\label{eq:deltInden}
\delta \left ( \vb{r} - \vb{r}_p (t) \right ) = \delta \left ( \vb{r} - \vb{r}_{0p}(t) \right ) \star_s \delta \left ( \vb{r} - \int_0^t d\tau \vb{v}_p (t) \right )
\end{equation}
Using this identity, it can be shown that (see Appendix)
\begin{equation}\label{eq:convContEq}
\partial_t \delta \left ( \vb{r} - \vb{r}_p (t) \right ) = - \div \left [ \vb{v}_p (t) \delta \left ( \vb{r} - \vb{r}_p (t) \right ) \right ] 
\end{equation}
The equation of continuity follows via
\begin{equation}\label{eq:particleChargeCont}
\begin{split}
-\partial_t \rho(t,\vb{r}) &=- q \sum_{p=1}^{N_p}\partial_t S(\vb{r}-\vb{r}_p(t)) =- q\sum_{p=1}^{N_p}S(\vb{r}) \star \partial_t \delta \left ( \vb{r}-\vb{r}_p(t) \right )\\
&=\div \left [ q\sum_{p=1}^{N_p} \vb{v}_p(t)S(\vb{r}- \vb{r}_p(t)) \right ]  =\div\mathbf{J}(t,\vb{r}).
\end{split}
\end{equation}

Of course, this is not the only identity that can be derived. As shown in the Appendix, another relationship  pertinent to this paper that  follows from the spectral representation of \eqref{eq:deltInden} is 
\begin{equation}\label{eq:chargeDef}
\rho(t, \mathbf{r})
= \rho(0,\vb{r}) - q \nabla \cdot \sum_{p = 1}^{N_p} \int_{\vb{r}_p(0)}^{\vb{r}_p(t)} d\tilde{\vb{r}}  S \left ( \mathbf{r} - \tilde{\mathbf{r} } (t) \right ) 
\end{equation}
As shown in the Appendix, this can be transformed into a line integral along the path taken by each particle.  Our analysis, thus far, has dwelt on the representation of the PSDF and its moments, viz., the charge and current. As is evident from \eqref{eq:particleChargeCont} that the equation of continuity is satisfied when the PSDF is represented by arbitrarily shaped particles. Next, we address mapping/measuring these charges and currents on grid (and how they interact with Maxwell's equations). 

\subsection{ Self-consistent solution of Maxwell's Equations with active sources}

For self consistency, one needs to solve Maxwell's equations with sources defined by \eqref{eq:sources}. In this pursuit, we discretize Maxwell's equations in space with a collection of  cells on which we define basis functions to represent fields and sources. Additionally, a temporal discretization is chosen such that a consistent solution for Newton's laws is obtained to update the position \emph{and} trajectory of the particle. 
It is well known that \emph{de-Rham} diagrams \cite{tonti01,deschamps81} provide appropriate map between fields, fluxes and sources in Maxwell's equations. In addition, one needs additional framework to ensure that charges and currents are correctly measured and integrated in time. In what follows, we discuss each in turn. 

\subsubsection{\emph{de-Rham} complex}

\begin{figure}
\centering
\includegraphics[scale=.3]{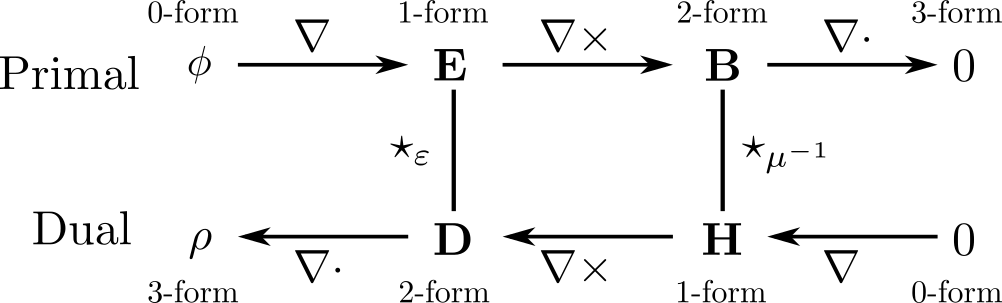}
\caption{Diagram of electromagnetic quantities as differential forms. }
\label{fig:primalANDdualSpaces}
\end{figure}

\begin{figure}
\centering
\includegraphics[scale =.5]{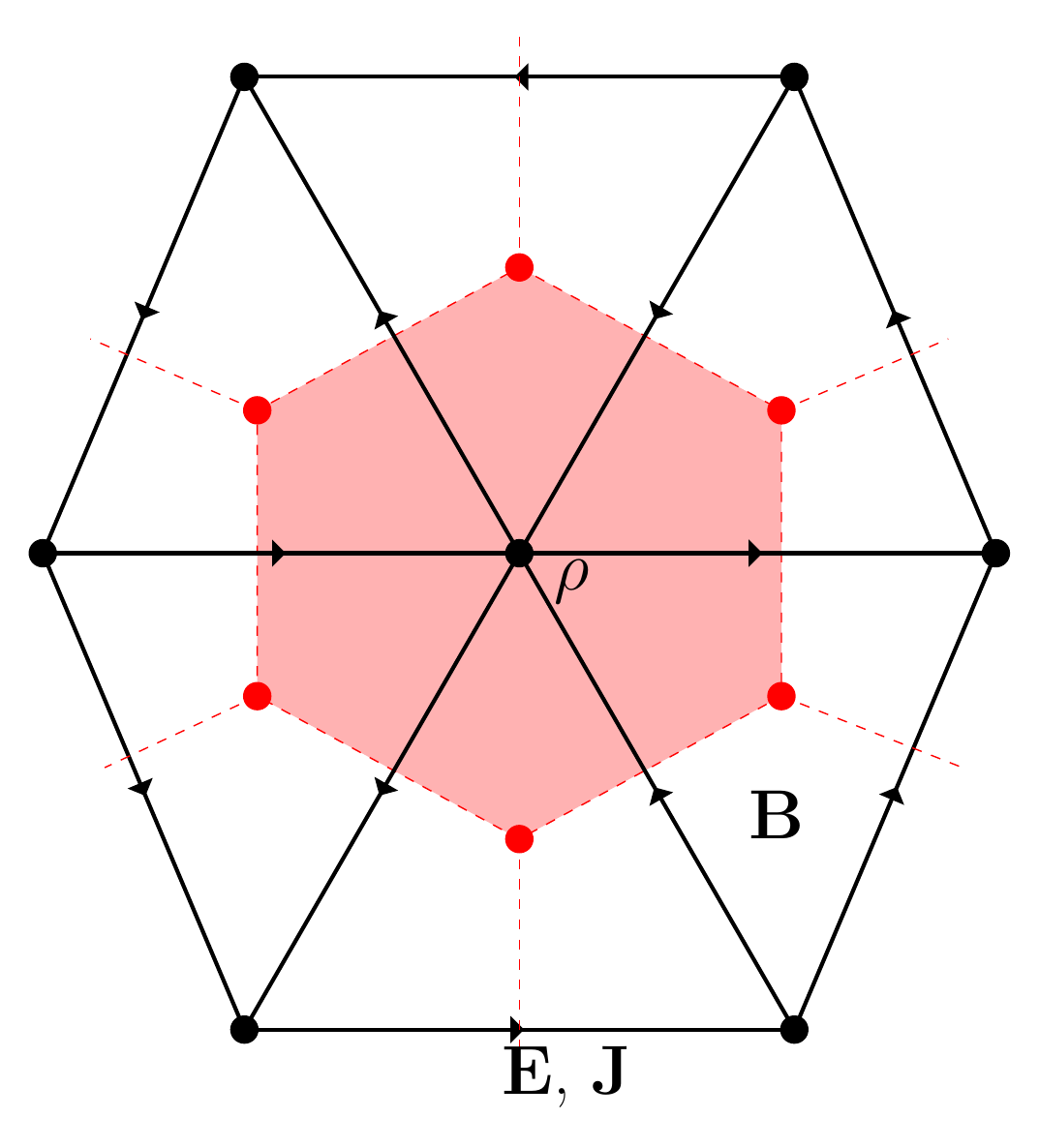}
\caption{Two dimensional diagram of where field and source quantities are defined on a simplical mesh. }
\label{fig:deRhamComplex}
\end{figure}

To begin, \emph{de-Rham} diagrams establish relations between field, fluxes and sources. These diagrams replicate Maxwell's equations in the discrete world; there exists extensive literature on their use in developing finite elements \cite{demk00,arnold10}. In a nutshell, these permit the definition of Whitney spaces that can be used to represent primary quantities ($\phi, \textbf{E},\textbf{B}$) or their duals ($\rho, \textbf{H},\textbf{D}$). It should be noted that the choice of fields on the primal or dual grids are not rigid; in other words, ($\rho, \textbf{H},\textbf{D}/\textbf{J}$) can be defined on the primal grid and ($\phi, \textbf{E},\textbf{B}$) on the dual grid. 
These relationships are shown in Figs. \ref{fig:primalANDdualSpaces} and \ref{fig:deRhamComplex}. Using Whitney spaces to solve Maxwell's equations has its antecedents in Bossavit's seminal work \cite{boss88} and those of several others \cite{rieben05,lee97,boss00,monk03,arnold06}; our presentation is along these lines. As has been shown in these papers, these basis sets and their function spaces are defined as follows: 
the 0-form defined on nodes, the 1-form defined on edges, the 2-form defined on faces, and the
3-form defined on volumes. In this work, we choose to use the Whitney basis sets that
\begin{subequations}\label{eq:hspaces}
\begin{equation}
\text{0-form} := W^{(0)} \in H^1
\end{equation}
\begin{equation}
\text{1-form} := \vb{W}^{(1)}(\vb{r}) \in H(curl)
\end{equation}
\begin{equation}
\text{2-form} := \vb{W}^{(2)}(\vb{r}) \in H(div)
\end{equation}
\begin{equation}
\text{3-form} := W^{(3)}(\vb{r}) \in L^2
\end{equation}
\end{subequations}
 The Hodge operator relates the corresponding field and flux quantities using the correct constitutive parameters, as well as provides a mapping between the primal and dual grids \cite{chew99,tarh99}.

\subsubsection{Discrete and self consistent Gauss' and Ampere's Laws}

In typical solutions to Maxwell equations,  there is an implicit assumption that the solution to the two curl equation enforces Gauss' laws together with the equation of continuity. While this is true in the continuous domain, one needs to handle things with care in the discrete domain. Specifically, the basis functions used to represent or measure the fields, fluxes, currents, and charges need to be consistent with each other. Let $\mathcal{L}_{\div} \circ $ denote a discrete divergence operator and $\mathcal{L}_{\curl} \circ $ denote a discrete curl operator, whose definitions are dependent on the basis functions. With no loss in generality, let's denote the basis functions, $\mathbf{N}_i (\mathbf{r})$ and $\xi_i(\vb{r})$, used to represent and measure the fields and fluxes, and charges, respectively. They belong to the appropriate space of functions depending on which quantity is being measured as determined by \eqref{eq:hspaces} and have local support over a simplex or cell. They also have relationships determined by the de-Rham complex shown in figure \ref{fig:primalANDdualSpaces}. With no loss of generality, let us define the vectors of quantities that are measured by our candidate basis functions. For instance, on can define
\begin{subequations}
\begin{equation}
d_i(t) = \langle \vb{N}_i(\vb{r}), \varepsilon \vb{E}(t,\vb{r})\rangle
\end{equation}
\begin{equation}
(\mathcal{L}_{\curl}\circ \vb{h}(t))_i = \langle \vb{N}_i(\vb{r}),\curl \mu^{-1}\vb{B}(t,\vb{r})\rangle
\end{equation}
\begin{equation}
    (\mathcal{L}_{\div} \circ
    \vb{d}(t))_i = \langle \xi_i(\vb{r}), \div \varepsilon\vb{E}(t,\vb{r}) \rangle
\end{equation}
\begin{equation}
j_i(t) = \langle \vb{N}_i(\vb{r}), \vb{J}(t,\vb{r}) \rangle
\end{equation}
\end{subequations}
Next, let us define $\vb{d}(t) = [d_1(t),d_2(t),\dots,d_{N_{\gamma}}(t)]$, $\vb{h}(t) = [h_1(t),h_2(t),\dots,h_{N_{\alpha}}(t)]$, and $\vb{j}(t) = [j_1(t),j_2(t),\dots,j_{N_{\beta}}(t)]$, which can be used to approximate the continuous functions, such as $\vb{D}(t,\vb{r}) = \sum_{i=1}^{N_\gamma} d_i(t)\vb{N}_i(\vb{r})$, where $N_\gamma$ is the number of degrees of freedom; as we will see, depending on the formulation it can either the number of edges or faces. This definition is the same for $N_{\alpha}$ and $N_{\beta}$. The degree of freedom vectors may require a Hodge operator to represent the function in the correct space.
For solutions to be consistent, the systems of equations should be such that the divergence of Ampere's law yields the time derivative of Gauss' law, which together with initial conditions will yield Gauss' law. While this is simply stated, there are a number of subtle nuances. In the discrete world, it follows that the same should be valid. Specifically, the spatial basis sets and the time integrator used has to be such that these relationships are satisfied. Note, these conditions must be satisfied exactly because only the curl equations, Faraday's law and Ampere's law, are explicitly solved. Gauss' law must be satisfied indirectly through the correct choice of spatial basis sets and time integration.

As our goal is to ensure the \emph{both} Gauss' laws and the equation of continuity are satisfied, it follows that there has to be a relationship between $\grad \xi_i (\vb{r})$ and $\mathbf{N}_i (\vb{r})$. We begin by considering the divergence of discrete Ampere's law;
specifically, 
\begin{equation}
\partial_t\mathcal{L}_{\div} \circ \vb{d}(t) =  \mathcal{L}_{\div} \circ \mathcal{L}_{\curl}\circ\vb{h}(t) - \mathcal{L}_{\div} \circ \vb{j}(t) 
\end{equation}
From the above, one can deduce the following: 
\begin{description}

\item [Condition 1: Consistent spatial basis sets] The basis sets should be such that the discrete divergence of the discrete rotation of the magnetic field is zero. In effect, 
the discrete differential operators should maintain the same properties as their continuous counterpart. To serve as the link between Ampere's law and the continuity equation, 
\begin{equation}
\label{eq:cond1}
\mathcal{L} _{\div} \circ \mathcal{L} _{\curl} \circ \vb{h}(t) = 0.
\end{equation} Though the discrete differential operators can be defined using properties of the mesh, they must be consistent with the discretized Faraday's and Gauss' law.

\item [Condition 2: Discrete divergence] Let $\xi (\vb{r})$ denote functions used to measure the charge on the mesh. Then $\grad \xi_i (\vb{r})$ should span the irrorational component of both $\vb{D}(t,\vb{r})$ and $\vb{J}(t,\vb{r})$. In effect, $\text{span} \left \{ \grad \xi_i (\vb{r}) \right \}  \subset \vb{D}(t,\vb{r}) \text{ or } \vb{J}(t,\vb{r})$ for $\forall t$. In other words, this representation is complete. The need for this condition arises from measuring both Gauss' law and the equation of continuity. Consider  measuring Gauss' law via
\begin{equation}
\label{eq:cond2}
\begin{split}
(\mathcal{L}_{\div}\circ \vb{d}(t))_i  &= \int_\Omega d\vb{r} \xi_i (\vb{r}) \rho (\vb{r})=  \int_\Omega d\vb{r}\xi_i(\vb{r}) \div \vb{D}(t,\vb{r}) \\
\int_\Omega d\vb{r} \xi_i (\vb{r}) \rho (\vb{r}) &= - \int_\Omega d\vb{r} \grad \xi_i (\vb{r}) \cdot \vb{D}(t,\vb{r}) + \int_{\partial \Omega} d\vb{r} \xi_i (\vb{r})\hat{n} \cdot \vb{D}(t,\vb{r}) .
\end{split}
\end{equation}
The preceding equation mathematically exemplifies this condition. An almost identical equation can be derived for the equation of continuity. It can be shown boundary terms in vanish interior to the domain and provided the correct boundary condition is imposed on $\hat{n} \cdot \vb{D}(t,\vb{r})$ on $\partial \Omega$, Gauss' law is implicitly satisfied together with condition 3.

\item [Condition 3: Discrete Time Integration] Finally, discrete time time stepping used to update the electric flux density in Ampere's equation and the path integral to evaluate the total current at any instant of time should be consistent with each other. This will ensure that the solution to Ampere's law agrees with Gauss' law. To gain more insight, we start by noting that when evolving Maxwell's equations over-time, we are in effect integrating the current; however, the time integral of the current is obtained via a path integral in \eqref{eq:chargeDef}. It follows, that if one were to take a discrete divergence of Ampere's laws, Condition 1 would yield
\begin{equation}
\label{eq:cond3}
 \partial_t \mathcal{L}_{\div} \circ \vb{d}(t) =  - \mathcal{L}_{\div} \circ \vb{j}(t)
\end{equation}
Provided Condition 2 is satisfied, it follows that choosing a consistent rule for integrating the charge along the particle path and integrating Ampere's law will ensure conservation of charge. 

\end{description}

\section{ Satisfaction of Conditions with Various EM-PIC Methods\label{sec:spatialBasis}}

Next, we examine three different formulations of EM-PIC; FDTD-PIC, EB-MFEM-PIC, and DH-MFEM-PIC in light of the conditions discussed thus far. While FDTD-PIC is well known, EM-MFEM-PIC and DH-MFEM-PIC place dominant importance on Faraday's and Ampere's laws, respectively. As a result, the dominant variables are represented on the primal grid, the others reside on dual grid. In what follows,  we seek to show how each formulation satisfies the three conditions. 

\subsection{FDTD-PIC \label{sec:FDTD}}

\begin{figure}
\centering
\includegraphics[scale=0.35]{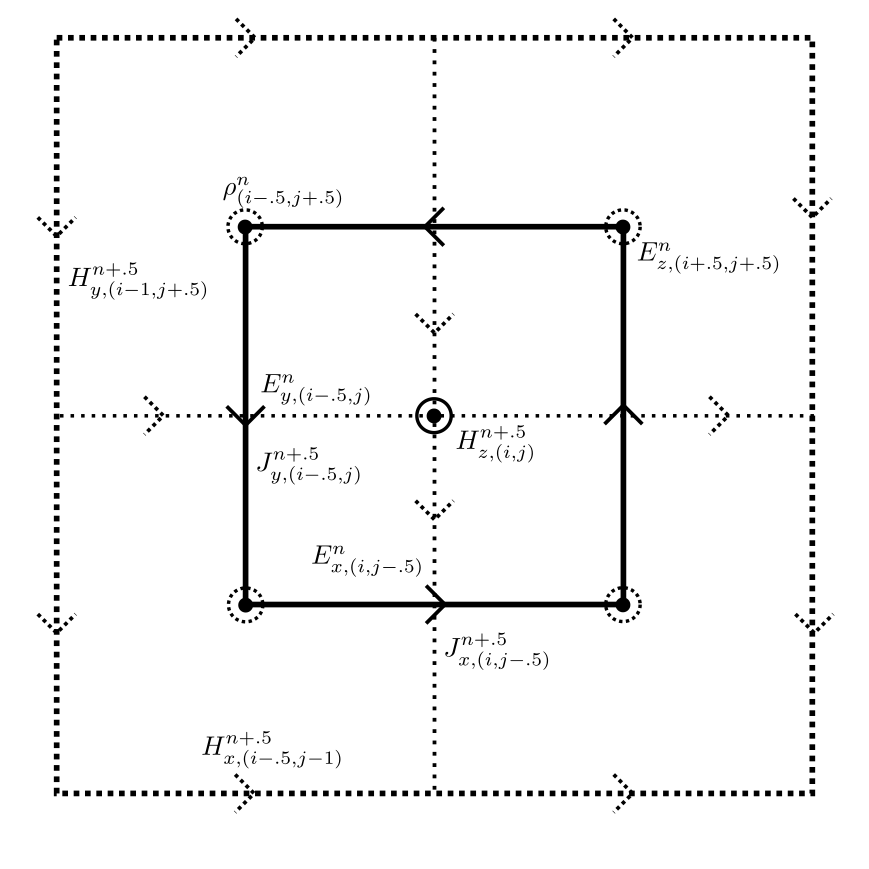}
\caption{FDTD 2D grid. Solid cell is the primal grid. Dotted cells are the dual grid.}
\label{fig:FDTDgrid}
\end{figure}

FDTD-PIC is the combination of the FDTD method \cite{yee} with PIC. It is discretized using staggered, structured grids in time and space, as seen in Figure \ref{fig:FDTDgrid} for two dimensions. For our discussion, the primal grid contains degrees of freedom for the electric field and magnetic flux density. Likewise, the dual grid has the degrees of freedom defined for the magnetic field and the electric flux density \cite{tonti01}.
Particles are mapped to the grid using the 3-forms that measure the charge density. In what follows, we will show that the particle weighting scheme defined by Langdon \cite{langdon} maps trivially onto the three conditions defined earlier.

\subsubsection{Satisfaction of Condition 1}
The discrete differential operators $\mathcal{L}_{\curl}\circ$ and $\mathcal{L}_{\div}\circ$ on the dual grid are defined as
\begin{equation}\label{eq:dcurl}
\mathcal{L}_{\curl}\circ = \begin{cases}
\pm 1 &\mbox{if } \hat{n}_j\cross\hat{l}_i \;\text{points into (out of) the face} \\
0 & \mbox{otherwise}
\end{cases}
\end{equation}
and
\begin{equation}\label{eq:ddiv}
\mathcal{L}_{\div}\circ = \begin{cases}
\pm 1 &\mbox{if } \hat{n}_j \text{ points out of (into) cell } c_i\\
0 & \mbox{otherwise}
\end{cases}
\end{equation}
The operator $\mathcal{L}_{\curl}{\circ} \in \mathbb{R}^{N_f \times N_e}$, where $N_f$ is the number of faces and $N_e$ is the number of edges. The operator $\mathcal{L}_{\div}\circ \in \mathbb{R}^{N_c \times N_f}$ where $N_c$ is the number of volumes.

\begin{figure}
\centering
\includegraphics[scale=0.55]{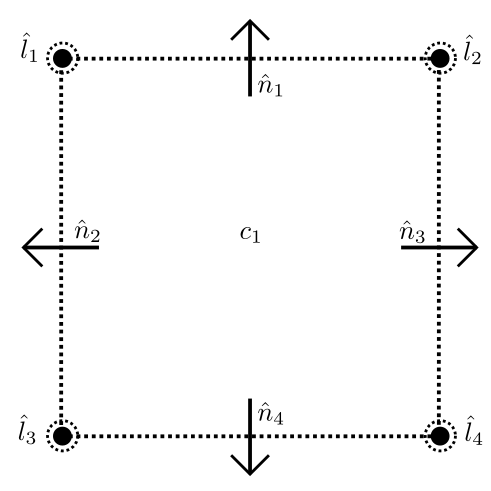}
\caption{Discrete curl and divergence on the dual grid. The curl is defined using the edges $\hat{l}_i$ that come out of the page. The divergence is defined using the face normals $\hat{n}_i$.}\label{fig:FDTDcurl}
\end{figure}

Consider an example mesh in Figure \ref{fig:FDTDcurl}. Assume the edge on the corners of the cell all point out of the page and the normal to the edges point out of the cell. The corresponding curl matrix definition would be
\begin{equation}
[\mathcal{L}_{\curl}\circ] = \begin{bmatrix}
1 & -1 & 0 & 0\\
-1 & 0 & 1 & 0\\
0 & 1 & 0 & -1\\
0 & 0 & -1 & 1
\end{bmatrix},
\end{equation}
The divergence matrix would be
\begin{equation}
[\mathcal{L}_{\div}\circ] = \begin{bmatrix}
1 & 1 & 1 & 1
\end{bmatrix}
\end{equation}
It is evident that in $\mathcal{L}_{\div}\circ\mathcal{L}_{\curl}\circ$, each entry in the resulting vector is zero, satisfying \eqref{eq:cond1}.

\subsubsection{Satisfaction of Condition 2}
The basis function $\xi(\vb{r})$ in \eqref{eq:cond2} is a 3-form, a constant in this formulation. Using the properties of 3-forms, \eqref{eq:cond2} reduces to
\begin{equation}\label{eq:cond2DHR}
\int_\Omega d\vb{r}\xi(\vb{r}) \div\vb{X}(\vb{r}) = \int_{\partial \Omega} d\vb{r} \hat{n} \cdot \vb{X} (\vb{r}) \xi(\vb{r}).
\end{equation} Consider the matrix form of $\mathcal{L}_{div}\circ$ based on Figure \ref{fig:FDTDcurl} 
\begin{equation}
\mathcal{L}_{\div}\circ = \begin{bmatrix}
1 & 0 & 1 & 1 & 0 & 1 & 0\\
0 & 1 & 0 & -1 & 1 & 0 & 1
\end{bmatrix}.
\end{equation}
A volume integral would be a summation over the cells of a mesh,
\begin{equation}
\sum_{i = 1}^{N_v} [\mathcal{L}_{\div}\circ]_{ij} = \begin{bmatrix}
1 & 1 & 1 & 0 & 1 & 1 & 1
\end{bmatrix}
\end{equation}
This would effect the surface integral in \eqref{eq:cond2DHR}. Therefore, the discrete continuity equation is satisfied exactly and consistency condition is fully satisfied.

\subsubsection{Satisfaction of Condition 3}
\begin{figure}
\centering
\includegraphics[scale=0.45]{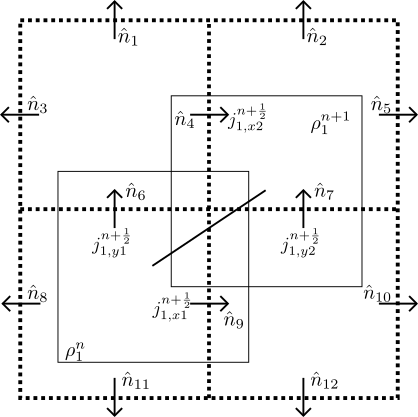}
\caption{FDTD 2D particle current mapping with particles of finite size. }
\label{fig:villPIC}
\end{figure}

Consider \eqref{eq:cond2DHR} in conjunction with \eqref{eq:cond3}. Using a leapfrog time marching scheme, the discrete continuity equation would be
\begin{equation}\label{eq:pathFDTD}
    [\mathcal{L}_{\div}]\frac{\vb{d}^{n+1}-\vb{d}^n}{\Delta_t} = -\frac{q}{\Delta_t}[\mathcal{L}_{\div}]\int_{\partial \Omega} d\vb{r}\int_{\vb{r}^n}^{\vb{r}^{n+1}}d\tilde{\vb{r}} \hat{n}\cdot \vb{p}(\tilde{\vb{r}}),
\end{equation}
where $\Delta_t$ is the time step size, $\vb{p}(\tilde{\vb{r}})$ is the path of the particle.  The integral computes the amount of charge that passes through a face.
 The divergence operator collects the contributions of each face, giving the net change of charge in each cell, as demonstrated in figure \ref{fig:villPIC}. This is equivalent to the current mapping in \cite{vill} for particles with finite shape. If the particle shape is infinitesimally small, it is equivalent to nearest-grid-point particle weighting.

\subsection{EB-MFEM-PIC Formulation \label{sec:EBFEM}}
Next, we analyze EB-MFEM-PIC as defined in \cite{boss88}. 
This finite element method can be seen as a generalization of FDTD. Like FDTD, it discretizes Maxwell's equations using a primal-dual grid system, with the field and flux quantities defined in the same manner. However, the grid itself can be unstructured, using non-cubic cells. The formulation for EB-MFEM can be found in appendix \ref{app:EB}. We have chosen to demonstrate the validity of the arguments in Section \ref{sec:framework} using brick elements as it is akin to FDTD-PIC. We defer the discussion on tetrahedral elements to the next section. 

\subsubsection{Satisfaction of Condition 1}
\begin{figure}
\centering
\includegraphics[scale=0.55]{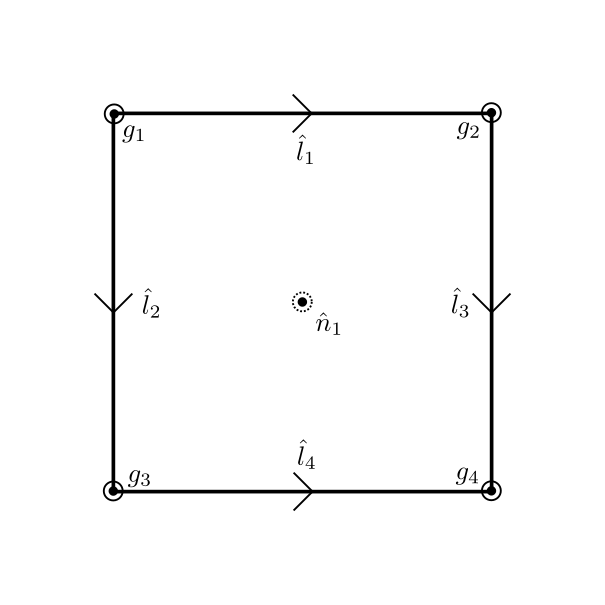}
\caption{FEM 2D curl and divergence mesh. The curl is defined using the edges $\hat{l}_i$ around face center $\hat{n}_i$. The divergence is defined with edges that point into node $g_i$.}
\label{fig:FEMcurl}
\end{figure}

The discrete differential operators $\mathcal{L}_{\curl}\circ$ and $\mathcal{L}_{\div}\circ$ are defined on the primal grid as
\begin{equation}
\mathcal{L}_{\curl}\circ = \begin{cases}
\pm 1 &\mbox{if } \hat{n}_j\cross\hat{l}_i \;\text{points into (out of) the face} \\
0 & \mbox{otherwise}
\end{cases}
\end{equation}
and
\begin{equation}
\mathcal{L}_{\div}\circ = \begin{cases}
\pm 1 &\mbox{if edge } j \text{ points into (out of) node } i\\
0 & \mbox{otherwise}
\end{cases}
\end{equation}
The operator $\mathcal{L}_{\curl}{\circ} \in \mathbb{R}^{N_e \times N_f}$. The operator $\mathcal{L}_{\div}\circ \in \mathbb{R}^{N_n \times N_e}$ where $N_n$ is the number of nodes. Consider the example mesh in Figure \ref{fig:FEMcurl}. The corresponding curl matrix definition would be
\begin{equation}\label{eq:FEMcurl}
\mathcal{L}_{\curl}\circ = \begin{bmatrix}
-1 & 1 & -1 & 1
\end{bmatrix}^T.
\end{equation}
The divergence matrix would be
\begin{equation}\label{eq:FEMdiv}
\mathcal{L}_{\div}\circ = \begin{bmatrix}
1 & 1 & 0 & 0\\
-1 & 0 & 1 & 0\\
0 & -1 & 0 & 1\\
0 & 0 & -1 & -1
\end{bmatrix}
\end{equation}
It is apparent that each entry in the resulting  $\mathcal{L}_{\div}\circ\mathcal{L}_{\curl}\circ$ is zero, satisfying \eqref{eq:cond1}.

\subsubsection{Satisfaction of Condition 2}

The basis function for the current density spans the gradient of the basis function for the charge density. The divergence operator $\mathcal{L}_{\div}\circ$ sums the edges connected to a node to satisfy \eqref{eq:cond2}. For perfect electrical conductor (PEC) boundary conditions, the current density normal to the face is zero, eliminating the surface integral term. This can be seen by summing the rows of \eqref{eq:FEMdiv}. For non-PEC boundaries, the domain is padded with ghost cells where the boundary nodes are not included. The non-terminating edges then make up the surface integral term when $\mathcal{L}_{\div}\circ$ is applied. Therefore, \eqref{eq:cond2} is satisfied.

\subsubsection{Satisfaction of Condition 3}
For each cell a particle visits, the corresponding current density is projected onto the edges of that cell.
Using \eqref{eq:cond2}, the discretized continuity equation that satisfies \eqref{eq:cond3} is
\begin{equation}\label{eq:pathFEM}
    [\mathcal{L}_{\div}]\frac{\vb{d}^{n+1}-\vb{d}^n}{\Delta_t} = -\frac{q}{\delta_t}[\mathcal{L}_{\div}]\int_{ \Omega} d\vb{r}\int_{\vb{r}^n}^{\vb{r}^{n+1}}d\tilde{\vb{r}} \grad W^0 \cdot \vb{p}(\tilde{\vb{r}}).
\end{equation}
This is equivalent to particle-in-cell area weighting in FDTD and the method used in \cite{east91,pintoMap}.
Due to the relationship between the 0-form on the primal grid and the 3-form on the dual grid, this is also equivalent to the approach defined in \cite{vill} for finite particle shapes.
\begin{figure}
\centering
\includegraphics[scale=0.45]{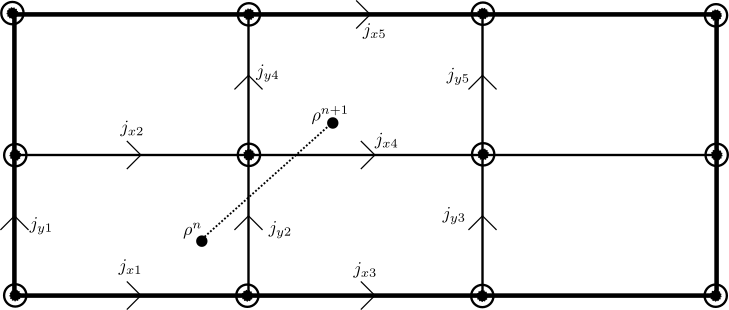}
\caption{FEM 2D particle current mapping with point particles. The particle current is mapped to each edge that encloses a cell along the particle path.}
\label{fig:FEMpic}
\end{figure}
\subsection{DH-MFEM-PIC Formulation \label{sec:DHFEM}}

The EB-MFEM-PIC formulation defined in \ref{sec:EBFEM} can be reformulated such that the primal grid has the degrees of freedom associated with the magnetic field and electric flux density. The dual grid contains the degrees of freedom for the electric field and magnetic flux density, still defined using the edges and faces, respectively.
The sources are represented on the primal grid using 3-forms for the charge density and 2-forms for the current density. 
The arguments for this formulation satisfying the three conditions are similar to that presented in section \ref{sec:FDTD}. Note, the principal ramification of this approach is that it is a natural framework for modeling sources as both the currents and electric flux density are modeled using the same function space, \emph{and} the fields, currents and charges lie in the primal grid. This is unlike EM-MFEM-PIC formulation where one needs transformations between primal and dual spaces. The ramification of these mappings become more consequential in higher order schemes--a topic that will be visited in a later paper.

\subsubsection{Satisfaction of Condition 1}
\begin{figure}
\centering
\includegraphics[scale=0.35]{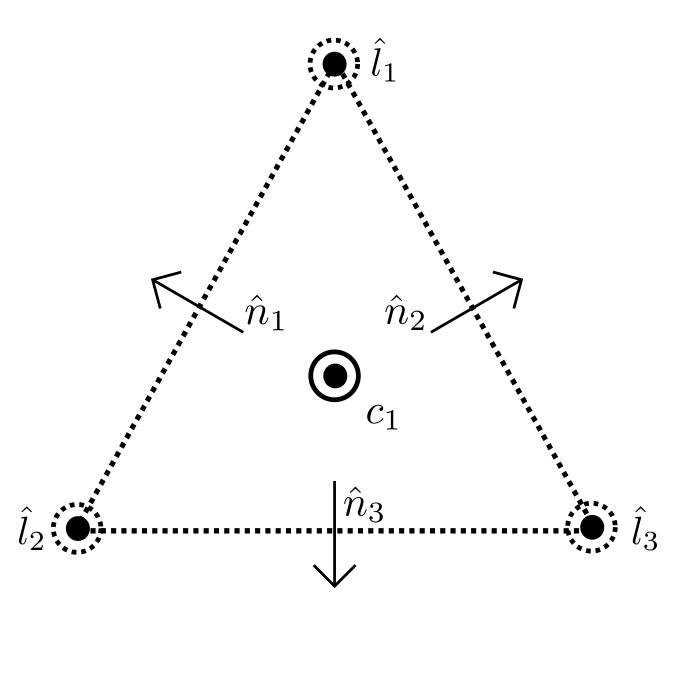}
\caption{Discrete curl mesh on the dual grid. The curl is defined with the outward pointing edge $\hat{l}_i$.}\label{fig:DHcurl}
\end{figure}

The discrete differential operators $\mathcal{L}_{\curl}\circ$ and $\mathcal{L}_{\div}\circ$ are defined on the primal grid, but have the same definitions as \eqref{eq:dcurl} and \eqref{eq:ddiv}. 
Consider the example mesh in Figure \ref{fig:DHcurl}. Assume the edge on the corners of the cell all point out of the page and the normal to the edges point out of the cell. The corresponding curl matrix definition for lowest order would be
\begin{equation}
[\mathcal{L}_{\curl}\circ] = \begin{bmatrix}
-1 & 1 & 0 \\
1 & 0 & -1 \\
0 & -1 & 1 
\end{bmatrix},
\end{equation}
The lowest order divergence matrix would be
\begin{equation}
[\mathcal{L}_{\div}\circ] = \begin{bmatrix}
1 & 1 & 1 
\end{bmatrix}
\end{equation}
Analyzing $\mathcal{L}_{\div}\circ\mathcal{L}_{\curl}\circ$, each entry in the resulting vector is zero, satisfying \eqref{eq:cond1}.

\subsubsection{Satisfaction of Condition 2}
 For lowest order, the continuity equation reduces to \eqref{eq:cond2DHR}.

\begin{figure}
\centering
\includegraphics[scale=0.45]{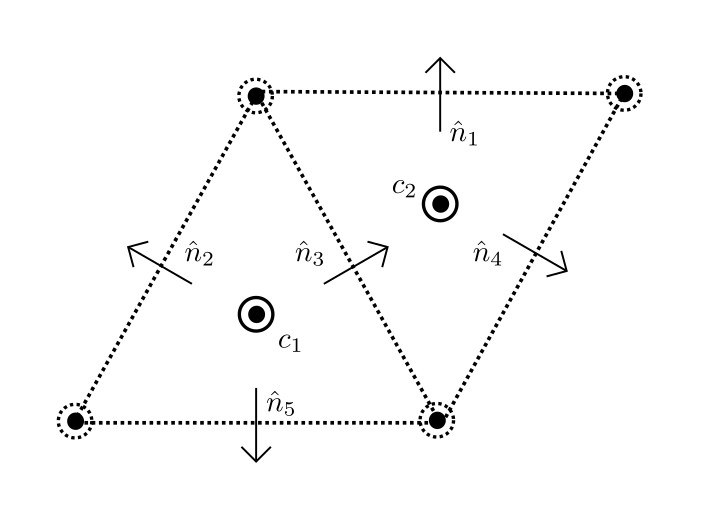}
\caption{FEM divergence mesh. The divergence is defined using face normals $\hat{n}_i$ that point out of cell $c_i$.}
\label{fig:DHdiv}
\end{figure}

Consider the matrix form of $\mathcal{L}_{\div}\circ$ based on figure \ref{fig:DHdiv} 

\begin{equation}
\mathcal{L}_{\div}\circ = \begin{bmatrix}
0 & 1 & 1 & 0 & 1 \\
1 & 0 & -1 & 1 & 0 
\end{bmatrix}.
\end{equation}
A volume integral would be a summation over the cells of a mesh,
\begin{equation}
\sum_{i = 1}^{N_v} [\mathcal{L}_{\div}\circ]_{ij} = \begin{bmatrix}
1 & 1 & 0 & 1 & 1
\end{bmatrix}
\end{equation}
This would effect the surface integral in \eqref{eq:cond2DHR}. Therefore, the discrete continuity equation is satisfied exactly and consistency condition is fully satisfied.
\subsubsection{Satisfaction of Condition 3}
\begin{figure}
\centering
\includegraphics[scale=0.35]{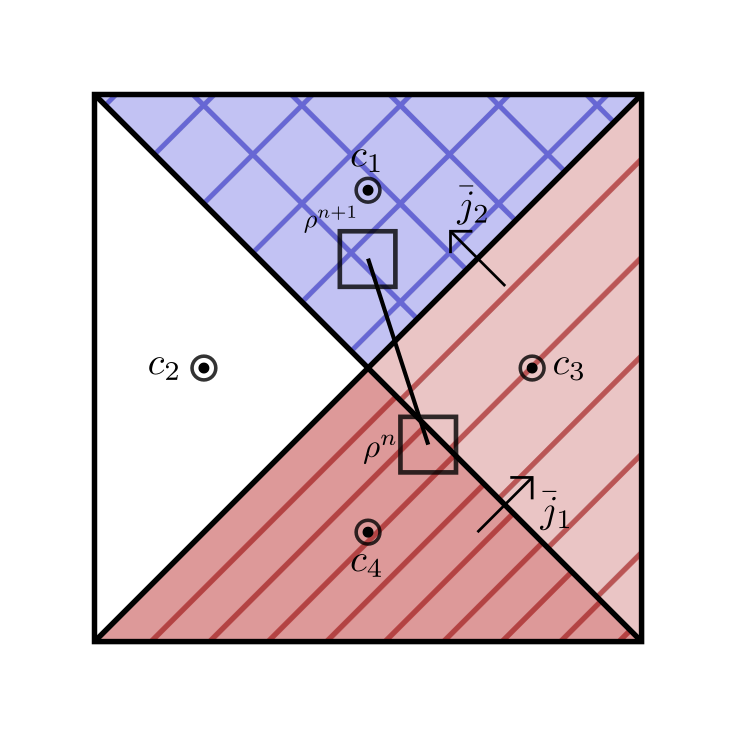}
\caption{FEM particle mapping with finite shape particles. The particle is mapped to cell $c_4$ and $c_3$ at first and ends in cell $c_1$ with current mapped at on the marked faces. The particle shape does not affect the conservation properties of this method. }
\label{fig:DHpart}
\end{figure}

Again, the satisfaction of condition three relies on the correct definition of current density on the faces of each cell.
For consistency of the change of charge density in cells, the amount of charge that passes through the face must be measured exactly.
This is equivalent to what is effectively done in \cite{whiteDH,tiexDH} through different techniques. Consider the mapping of current defined by the particle in Figure \ref{fig:DHpart}. Here, though the particle moves at an angle with respect to the face, the current density as seen by the mesh moves in the direction of the face normal. The solver in effect sees a charge density move from cell $c_4$ to $c_3$ and the total charge density in $c_3$ moves to $c_1$ by time step $n+1$.
\section{Consequences of Inconsistency \label{sec:inconsistent}}

In the previous sections, the three rule framework was defined and applied to different formulations of EM-PIC.
The references to previous works show how these rules encompasses well known EM-PIC formulations. Next, we will examine ramifications of violating these three rules. By examining potential pitfalls and how it manifests in results, one can more easily diagnose errors in new EM-PIC schemes. 

\subsection{ Violation of Condition 1}

First, consider violations of the first condition.
Failing to satisfy \eqref{eq:cond1} can arise from choosing the wrong basis sets to discretize \eqref{eq:maxwell}. 
One way to do so  is to choose basis sets that do not obey the de-Rahm complex. 
For example,  if one solves both Ampere's law and Faraday's law on the same grid, staggered in time. 
Like in FDTD, we consider representing the electric field and magnetic field; 
however, this time it is on the same grid. Assume the fields are represented by 1-forms and test \eqref{eq:far} and \eqref{eq:amp} with the same basis functions. 
The discretized system is
\begin{subequations}
\begin{equation}
\vb{h}^{n+\frac{1}{2}} =\vb{h}^{n+\frac{1}{2}} - \Delta_t[\mathcal{L}_{\curl}'\circ]\vb{e}^n
\end{equation}
\begin{equation}
\vb{e}^{n+1} =\vb{e}^{n+1} + \Delta_t([\mathcal{L}_{\curl}'\circ]\vb{h}^{n+\frac{1}{2}} - \vb{j}^{n+\frac{1}{2}}),
\end{equation}
\end{subequations}
where $\mathcal{L}_{\curl}'\circ$ is a modified curl operator, as the previously defined operator has the wrong dimensions. 
If this scheme is done with a structured mesh in FDTD, one would have a higher error as the curl would be represented using forward or backward difference rather than a central difference. 
Instead of the recurrence scheme found in \eqref{eq:FDTDampt}, one would have
\begin{equation}
\begin{split}
E_{x,(i+\frac{1}{2},j,k)}^{n+1} = E_{x,(i+\frac{1}{2},j,k)}^{n}& +\frac{\Delta_t}{\varepsilon\Delta_y}(H_{z,(i+\frac{1}{2},j+1,k)}^{n+\frac{1}{2}} -H_{z,(i+\frac{1}{2},j,k)}^{n+\frac{1}{2}})\\
&-\frac{\Delta_t}{\varepsilon\Delta_z}(H_{y,(i+\frac{1}{2},j,k+1)}^{n+\frac{1}{2}} -H_{y,(i+\frac{1}{2},j,k)}^{n+\frac{1}{2}})\\
&- \frac{\Delta_t}{\varepsilon}J_{x,(i+\frac{1}{2},j,k)}^{n+1/2}
\end{split}
\end{equation}
if a forward difference was used. 

The effects of this are more clear from an FEM perspective. 
Consider trying to find modes of a perfect electric conductor rectangular cavity using the generalized eigenvalue problem formed by the time harmonic form of \eqref{eq:genEig}
\begin{equation}\label{eq:genEig}
\begin{bmatrix}
0 & [\mathcal{L}^*_{\curl}\circ] \\
-[\mathcal{L}^*_{\curl}\circ]^T[\star_{\mu^{-1}}] & 0 
\end{bmatrix}\begin{bmatrix}
\vb{b}\\\vb{e}
\end{bmatrix}=
\Lambda\begin{bmatrix}
[\vb{I}] & 0 \\
0 & [\star_\varepsilon] 
\end{bmatrix}\begin{bmatrix}
\vb{b}\\\vb{e}
\end{bmatrix}
\end{equation}
where the eigenvalues in $\Lambda = j\omega$. 
The dimensions of the cavity are \SI{.6}{m}$\times$\SI{.4}{m}$\times$\SI{1}{m}.
In a consistent scheme, $[\mathcal{L}^*_{\curl}\circ] = [\mathcal{L}_{\curl}\circ]$ and $[\star_{\mu^{-1}}]$ is as defined in \eqref{eq:genEig}. 
However, in the inconsistent scheme we have defined,
\begin{equation}
[\mathcal{L}^*_{\curl}\circ]_{i,j} = \int d\Omega \vb{W}^{(1)}_{i}\cdot\curl\vb{W}^{(1)}_j   
\end{equation}
and $[\star_{\mu^{-1}}] = [\star_\varepsilon]$. 
When the inconsistent scheme is used, spurious modes are found that do not correspond to the true modes of the cavity as seen in table \ref{tab:cavModes}.
Using this inconsistent scheme could permit incorrect solutions into the field solver, invalidating results.

\begin{table}[]
\centering
\begin{tabular}{|c c c |}
\hline
$f_{\text{true}}$ & $f_{\text{consistent}}$ & $f_{\text{inconsistent}}$ \\
\hline
291.346 & 290.829 & 0.280 \\
390.242 & 389.157 & 1.036 \\
403.608 & 402.697 & 1.841 \\
450.382 & 449.654 & 2.451 \\
\hline
\end{tabular}
\caption{First four rectangular cavity modes for consistent and inconsistent FEM solver in MHz }    \label{tab:cavModes}    
\end{table}

\subsection{Violation of Condition 2}
Next, we consider the repercussions of violating Condition 2.
Consider the linear mapping scheme typically found in early PIC literature where the weighting function is defined as
\begin{equation}
W\left(\vb{r}-\vb{r}_p(t)\right) = \prod_{n\in\{x,y,z\}} 1-\frac{|r_{n,g}-r_{n,p}|}{\Delta_n}.
\end{equation}
When the particle is mapped to the nodes as the charge density, $r_{n,g}$ is the $x,y,z-$ coordinate of the node. When mapped to the edges as the current density, $r_{n,g}$ is the $x,y,z-$ coordinate of the center of the edge. 
It is well known that this mapping does not satisfy the continuity equation. 
For example, consider the particle motion in Figure \ref{fig:FEMpic} where the particle moves from $(.3,.4)$ to $(.7,.6)$. 
Table \ref{tab:badMap} shows the error in the continuity equation using this mapping. 
The result is an inconsistency in the solved Ampere's law and measured Gauss' law due to spurious charge seen by the field solver as the next mapping is done. 
This mapping scheme does not appropriately satisfy \eqref{eq:cond2}, as the combination of basis functions for the current density at the node is not exactly equal to the gradient of the charge density basis function. Despite being in the same direction, the basis function defined by the linear mapping for the current density is different than the basis functions used to measure the fields. Therefore, the de-Rham complex is not truly satisfied and the spatial link between Gauss' law and Ampere's law is no longer preserved.
\begin{table}
\centering
\begin{tabular}{|c | c c c |}
\hline
&  $\partial_t \rho$ & $ \div \vb{J}$ & $\partial \rho + \div \vb{J}$\\
\hline
node 1 & -0.100 & -0.050 & -0.150 \\
node 2 &  0.300 & -0.850 & -0.550 \\
node 3 & -0.300 &  0.850 &  0.550 \\
node 4 &  0.100 &  0.050 &  0.150 \\
\hline
\end{tabular}
\caption{Continuity equation error for linear mapping }    \label{tab:badMap}    
\end{table}

\subsection{Violation of Condition 3}
Next, we examine  how failing to correctly integrate the spatial integral in \eqref{eq:cond3} results in a violation of Gauss' law and the continuity equation due to spurious charge on the grid.
Consider the movement of a particle on a linear path that remains in a single cell using a 0-form basis set to measure the charge density. 
For a linear path, the path integral can be exactly computed either using an analytic scheme \cite{moon} or an one point integration rule, as seen in Table \ref{tab:singIErr}. 
This translates to machine precision error in the continuity equation in Table \ref{tab:singCE}.
\begin{table}
\centering
\begin{tabular}{|c | c c c |} 
\hline
& Analytic & 1 pt Quad & Error \\ 
\hline
Edge 1 & 1.248 & 1.248 & -4.441e-16 \\ 
Edge 2 & 0.505 & 0.505 &  0.000 \\
Edge 3 & 0.248 & 0.248 &  0.000 \\
\hline
\end{tabular}

\caption{Error of Analytic Integration vs Quadrature Rule for Linear Path}\label{tab:singIErr}
\end{table}      
\begin{table}
\centering
\begin{tabular}{|c | c c c c c|} 
\hline
& $Q^{n+1}$ & $Q^{n}$ & $\frac{ Q^{n+1}-Q^{n}}{\Delta_t}$ &  $\nabla \cdot \vb{J}$ & $ \frac{ Q^{n+1}-Q^{n}}{\Delta_t} + \nabla \cdot \vb{J}$ \\ 
\hline
Node 1 & 0.284 & 0.459 & -1.753 &  1.753 &  0.000 \\ 
Node 2 & 0.550 & 0.450 &  1.000 & -1.000 & -4.440e-16 \\
Node 3 & 0.166 & 0.091 &  0.753 & -0.753 &  0.000 \\
\hline
\end{tabular}
\caption{Continuity Equation Error for Linear Path}\label{tab:singCE}
\end{table}

Similarly, consider the case when a particle crosses at least one face on its path. 
When the path is in multiple cells, an error occurs when the path integral is not calculated in each cell, as seen in Table \ref{tab:mcCE}. 
This is an inconsistency in \eqref{eq:cond3} as though the time integral may be computed correctly, parts of the path integral are missing.

\begin{table}
\centering
\begin{tabular}{|c | c c |}
\hline
&  $ \frac{ Q^{n+1}-Q^{n}}{\Delta_t} + \nabla \cdot \vb{j}^{n+1/2}_{nc}$ & $ \frac{ Q^{n+1}-Q^{n+\alpha}}{\Delta_{t2}} +\frac{ Q^{n+\alpha}-Q^{n}}{\Delta_{t1}}  + \nabla \cdot \vb{j}^{n+1/2}_{cc}$\\
\hline
Node 1 &  1.008 &  0.000 \\
Node 2 &  0.394 & -4.441e-16 \\
Node 3 & -0.008 &  2.220e-16 \\
Node 4 & -1.394 & -4.441e-16 \\
\hline
\end{tabular}
\caption{Continuity Equation Error for Multiple Cells}    \label{tab:mcCE}    
\end{table}

\begin{table}
\centering
\begin{tabular}{|c | c |c  c | c c|}
\hline
& Analytic & 1 pt Quad & 1 pt error & 2 pt Quad & 2 pt error\\ 
\hline
Edge 1 & 1.253 & 1.250 & -0.003 & 1.253 & -6.661e-16 \\ 
Edge 2 & 0.500 & 0.500 &  0.000 & 0.500 &  2.220e-16  \\
Edge 3 & 0.253 & 0.250 &  0.003 & 0.253 &  0.000 \\
\hline
\end{tabular}
\caption{Error of Analytic Integration vs Quadrature Rule for Quadratic Path} \label{tab:quadIErr}
\end{table}

\subsection{Results for Consistent DH-MFEM-PIC}

\begin{figure}
\begin{subfigure}[h]{.5\textwidth}
\centering
\includegraphics[scale=0.48]{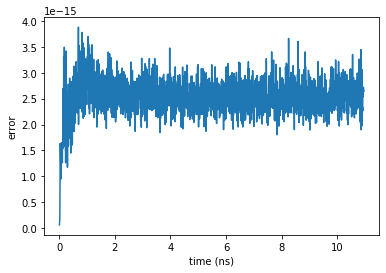}
\caption{ }
\label{fig:DHgcontE}
\end{subfigure}
\begin{subfigure}[h]{.5\textwidth}
\centering
\includegraphics[scale=0.48]{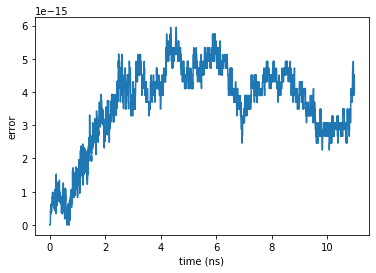}
\caption{ }
\label{fig:DHggausE}
\end{subfigure}
\caption{Error in the continuity equation (\ref{fig:DHgcontE}) and Gauss' law (\ref{fig:DHggausE}) for the particle beam case. }\label{fig:DHerrs}
\end{figure}

\begin{figure}
\centering
\includegraphics[scale=0.45]{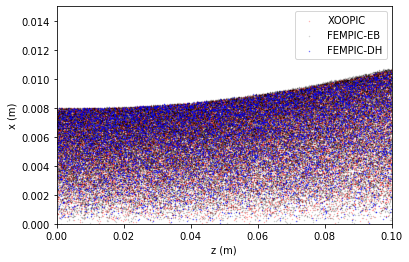}
\caption{Particle Beam for FDTD, $EB$, and $DH$ formulation }
\label{fig:EBDHbeam}
\end{figure}

Finally, we examine results when all three conditions are satisfied. We have chosen to illustrate these using DH-MFEM-PIC as results for EB-MFEM-PIC are available \cite{pintoMap,moon,oconnor}. First, we consider a particle beam in a 0.1 m PEC cylindrical cavity with a radius of 0.02 m. The particle beam has a voltage of 7.107 kV and current of 0.25A, simulated by inserting ten particles per time step. The starting velocity of the particles entering the tube is $5\times 10^4$ km/s with a beam width of 0.008 m.
In this test, when the three conditions are satisfied, the beam smoothly spreads from a uniform distributions.
Using a DH-based EM-PIC formulation and point particles for particle shape, by following section \ref{sec:DHFEM}, a self-consistent charge conserving EM-PIC scheme was created. 
Point particles are used because without introducing errors, it removes the need to compute the volume of a particle passing through a face, which can be extremely difficult in three dimensions,
The error in the continuity equation and Gauss law is found in Figure \ref{fig:DHerrs}, showing that satisfying conditions two and three is possible with a nearest grid point algorithm in FEM. 
The trend of the spreading beam is compared to that of an EB-MFEM-PIC and a FDTD based EM-PIC formulation in Figure \ref{fig:EBDHbeam}.

\begin{figure}
\centering
\includegraphics[scale=0.45]{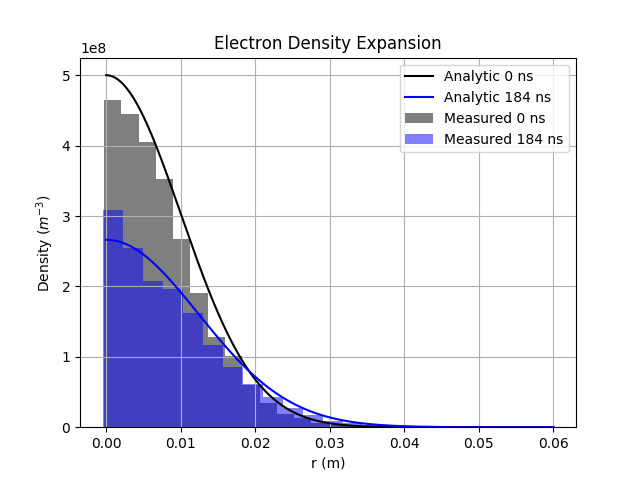}
\caption{Electron density of expanding adiabatic plasma ball using $\vb{DH}$-MFEM-PIC compared to analytic expected densities.}\label{fig:DHpball}
\end{figure}

Lastly, we examine the expansion of a plasma ball.
A spherical plasma ball is given a Gaussian distribution of ions and electrons in a geometry with first order absorbing boundary conditions. The experiment was simulated with the DH-FEM formulation with a point particle shape. A combination of $Sr^+$ ions at 1K and electrons at 100K are placed at a density of $5\times 10^8$ particles per cubic meter such that the system is initially charge neutral.
Further details on the experiment setup can be found in \cite{oconnor} and derivations of the analytic results in \cite{kovalev}. 
As is evident from Figure \ref{fig:DHpball}, we achieve excellent agreement with the analytic results. Again, because all the conditions are satisfied, the continuity equation and errors in Gauss' law are kept to machine precision.

\begin{figure}
\centering
\includegraphics[scale=0.45]{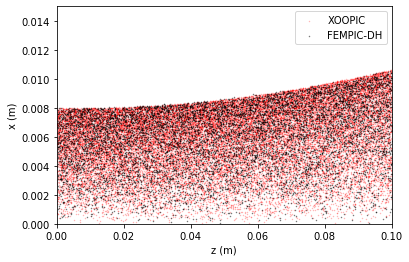}
\caption{Particle beam comparison of $\vb{DH}$-MFEM-PIC and FDTD-PIC.}\label{fig:DHgbeam}
\end{figure}

\section{Summary and conclusions}

In this work, we have presented a  framework to define self-consistent EM-PIC schemes. By considering the space of functions used to represent fields, fluxes, and sources, one can ensure that all the relationships between Maxwell's equations are preserved in a consistent manner. This has been demonstrated for FDTD and two formulations of FEM, and shown how it generalizes and relates to well known previous work. The framework is easily related to other field solvers not mentioned directly provided one considers the relationship of the basis functions. Future work will apply this framework to higher order basis functions and particle paths\cite{oconnorTime}. 

\section{Acknowledgements}

This work was sponsored by the US Air Force Research Laboratory under contracts FA8650-19-F-1747 and FA8650-20-C-1132. This work was also supported by the Department of Energy Computational Science Graduate fellowship under grant DE-FG02-97ER25308. The authors would also like to thank the HPCC Facility at
Michigan State University and the DoD SMART scholarship.

\section{Data Availability} 
The data that support the findings of this study are available from the corresponding author
upon reasonable request.

\section{Appendix}

\subsection{Calculus on Distribution functions}

 The following must be understood in the sense of distributions. Specifically, we consider derivatives and integral of the delta function. These appear in \eqref{eq:samplingPhase} onwards where one represents the spatial variation of PSDF using $S(\vb{r} - \vb{r}_p(t)) = S(\vb{r}) \star_s \delta \left (\vb{r} - \vb{r}_p(t) \right )$. Calculus on delta function is critical for developing insight into implementing equations of continuity and Gauss' law on the grid. To do so, we start with a transform to $\vb{k}$-space to prove \eqref{eq:convContEq} and \eqref{eq:chargeDef}.

 \subsubsection{Time derivatives}
 
 Using $\vb{r}_p(t) = \vb{r}_{0p} + \int_0^t d\tau \vb{v}_p (\tau)$, the time derivative of $\delta \left (\vb{r} - \vb{r}_p(t) \right )$  can be written as  
\begin{equation}
\partial_t\delta\left (\vb{r}-\vb{r}_p(t)\right )= \delta(\vb{r}-\vb{r}_{0p}) \star_s\partial_t \delta\left (\vb{r}- \int_0^t d\tau\vb{v}_p(\tau)\right )
\end{equation}
Next, we exploit 
\begin{equation}\label{eq:deltdef}
\delta(\vb{r} - \vb{a}) = \frac{1}{(2\pi)^3}\int_{-\infty}^{\infty}d^3\vb{k} e^{j\vb{k}\cdot(\vb{r}-\vb{a})}
\end{equation}
to yield
\begin{equation}
\begin{split}
\partial_t \delta \left (\vb{r} - \int_0^t d\tau\vb{v}_p(\tau) \right ) &=\frac{1}{(2\pi)^3}\int_{-\infty}^{\infty}d^3\vb{k} \partial_t e^{j\vb{k}\cdot\left(\vb{r}-\int_0^t d\tau\vb{v}_p(\tau)\right)}\\
&=-\div\vb{v}_p(t)\delta\left (\vb{r}-\int_0^t d\tau\vb{v}_p(\tau)\right ).
\end{split}
\end{equation}
Therefore
\begin{equation}
\begin{split}
\partial_t\delta(\vb{r}-\vb{r}_p(t)) &= -\delta(\vb{r}-\vb{r}_{0p}) \star_s\div\vb{v}_p(t)\delta\left (\vb{r}-\int_0^t d\tau\vb{v}_p(t)\right )\\
&=-\div \left [\vb{v}_p(t)\delta\left (\vb{r}-\vb{r}_p(t)\right ) \right ]
\end{split}
\end{equation}

\subsection{Time integrals}

Next, starting from the time integral of \eqref{eq:convContEq} and using the definition of current density in equation \eqref{eq:Jsource},
\begin{equation}\label{eq:cdStart}
\begin{split}
\int_0^t d\tau \div \vb{v}_p(\tau)&S(\vb{r}- \vb{r}_p(\tau))\\
&=S(\vb{r})\star_s \int_0^t d\tau \div\vb{v}_p(\tau) \delta\left (\vb{r} - \int_0^\tau d\tau'\vb{v}_p(\tau')\right )
\end{split}
\end{equation}
Again, using \eqref{eq:deltdef} and $\partial_t \tilde{\vb{r}}_p(t) = \vb{v}_p (t)$,
\begin{equation}
\begin{split}
    & \int_0^t d\tau \div\vb{v}_p(\tau) \delta\left (\vb{r} - \int_0^\tau d\tau'\vb{v}_p(\tau')\right ) \\
    & 
    = \int_0^t \frac{d\tau}{(2\pi)^3} \div\int_{-\infty}^\infty d^3\vb{k} \vb{v}_p(\tau) e^{j\vb{k}\cdot\left(\vb{r} - \int_0^{\tau} d\tau' \vb{v}_p(\tau')\right)}\\
&= \div \frac{1}{(2\pi)^3}\int_{-\infty}^\infty d^3\vb{k} e^{j\vb{k}\cdot\vb{r}} \int_0^t d\tau \vb{v}_p(\tau)e^{-j\vb{k}\cdot\int_0^\tau d\tau' \vb{v}_p(\tau')} \\
& = \div \frac{1}{(2\pi)^3}\int_{-\infty}^\infty d^3\vb{k} e^{j\vb{k}\cdot\vb{r}} \int_{\vb{r}_p(0)}^{\vb{r}_p(t)} d\tilde{\vb{r}}_p e^{-j\vb{k}\cdot\tilde{\vb{r}}_p} 
\end{split}
\end{equation}
Interchanging the order of the integrals in the above equation results in 
\begin{equation}
 \int_0^t d\tau \div \vb{v}_p(\tau)S(\vb{r}- \vb{r}_p(\tau)) = \div\int_{\vb{r}_p(0)}^{\vb{r}_p(t)} d\tilde{\vb{r}} S\left(\vb{r} -\tilde{\vb{r}} \right)
\end{equation}
This equation can be used to define the charge density in terms of the particle's path; it can also be used to define the time integral of the current density. 

\subsection{Maxwell Solvers}

The following sections are presented purely for completeness and to conform with material presented earlier. We present details of the three Maxwell solvers used here; FDTD, EB-MFEM, and DH-MFEM.

\subsubsection{FDTD}

Define the rooftop function $\Lambda_i(\vb{r})$ 
\begin{equation}
\Lambda_i(\vb{r}) = \begin{cases}
1 - \frac{|\hat{i}\cdot\vb{r}|}{\Delta_i} &  0 < |\hat{i}\cdot\vb{r}| \le \Delta_i \\
0 & \text{o.w.}
\end{cases}
\end{equation}
 where $i = x,y,z$. The basis functions to discretize Maxwell's equations for FDTD are
\begin{subequations}
	\begin{equation}
\text{1-form} := \vb{u}_i\Lambda_{i+1}(\vb{r}-\vb{l}_m)\Lambda_{i+2}(\vb{r}-\vb{l}_m)
	\end{equation}
	\begin{equation}
\text{2-form} := \vb{u}_i\Lambda_{i}(\vb{r}-\vb{f}_m)
	\end{equation}
\end{subequations}
with $\vb{u}_i = \hat{x},\hat{y},\hat{z}$, $\vb{l}_m$ is the midpoint of each edge, and $\vb{f}_m$ is the midpoint of each face. 
Testing Faraday's law with the normal to cell faces and Ampere's law is with the edge unit vectors yields
\begin{subequations}
\begin{equation}\label{eq:FDTDfart}
\begin{split}
\langle \vb{u}_i\delta((\vb{r}-\vb{f}_m)\cdot\hat{i}),\partial_t \mu\vb{H}(t,\vb{r}) \rangle &= - \langle \vb{u}_i\delta((\vb{r}-\vb{f}_m)\cdot\hat{i} ),\curl \vb{E}(t,\vb{r}) \rangle\\
\partial_t\vb{H}_i(t,\vb{f}_m) &= -\mu_0^{-1}\vb{u}_i\cdot\curl\vb{E}(t,\vb{f}_m)\\
&=\mu_0^{-1}\partial_{i+1}\vb{E}_{i+2}(t,\vb{f}_m)\\
&-\mu_0^{-1}\partial_{i+2}\vb{E}_{i+1}(t,\vb{f}_m)\\
\end{split}
\end{equation}
\begin{equation}\label{eq:FDTDampt}
\begin{split}
\langle &\vb{u}_i\delta((\vb{r}-\vb{f}_m)\cdot\hat{i}),\partial_t \varepsilon\vb{E}(t,\vb{r}) \rangle = \\
&- \langle \vb{u}_i\delta((\vb{r}-\vb{l}_m)\cdot\hat{i} ),\curl \vb{H}(t,\vb{r}) \rangle
-\langle \vb{u}_i\delta((\vb{r}-\vb{l}_m)\cdot\hat{i} ),\vb{J}(t,\vb{r}) \rangle\\
&\partial_t\vb{E}_i(\vb{l}_m) = -\varepsilon_0^{-1}\vb{u}_i\cdot\curl\vb{H}(t,\vb{l}_m)-\varepsilon_0^{-1}\vb{u}_i\cdot\vb{J}(t,\vb{l}_m)\\
&=\varepsilon_0^{-1}\partial_{i+1}\vb{H}_{i+2}(t,\vb{l}_m)-\varepsilon_0^{-1}\partial_{i+2}\vb{H}_{i+1}(t,\vb{l}_m)-\varepsilon_0^{-1}\vb{J}_{i}(t,\vb{l}_m).\\
\end{split}
\end{equation}
\end{subequations}
 The addition on index $i$ follows cycle $\{x,y,z\}$.
This testing is equivalent to the usual presentation of the FDTD recurrence relation. For example, consider $i=x$ in \eqref{eq:FDTDfart} would yield
\begin{equation}
\begin{split}
H_{x,(i,j+\frac{1}{2},k+\frac{1}{2})}^{n+\frac{1}{2}} = H_{x,(i,j+\frac{1}{2},k+\frac{1}{2})}^{n-\frac{1}{2}}& -\frac{\Delta_t}{\mu\Delta_y}(E_{z,(i,j+1,k+\frac{1}{2})}^n -E_{z,(i,j,k+\frac{1}{2})}^n)\\
&+\frac{\Delta_t}{\mu\Delta_z}(E_{y,(i,j+\frac{1}{2},k+1)}^n -E_{y,(i,j+\frac{1}{2},k)}^n).
\end{split}
\end{equation}
Likewise, $i=x$ in \eqref{eq:FDTDampt} would yield
\begin{equation}
\begin{split}
E_{x,(i+\frac{1}{2},j,k)}^{n+1} = E_{x,(i+\frac{1}{2},j,k)}^{n}& +\frac{\Delta_t}{\varepsilon\Delta_y}(H_{z,(i+\frac{1}{2},j+\frac{1}{2},k)}^{n+\frac{1}{2}} -H_{z,(i+\frac{1}{2},j-\frac{1}{2},k)}^{n+\frac{1}{2}})\\
&-\frac{\Delta_t}{\varepsilon\Delta_z}(H_{y,(i+\frac{1}{2},j,k+\frac{1}{2})}^{n+\frac{1}{2}} -H_{y,(i+\frac{1}{2},j,k-\frac{1}{2})}^{n+\frac{1}{2}})\\
&- \frac{\Delta_t}{\varepsilon}J_{x,(i+\frac{1}{2},j,k)}^{n+1/2}.
\end{split}
\end{equation}

\subsubsection{EB-MFEM Formulation\label{app:EB}}

In this formulation, $\vb{E}(t,\vb{r}) \approx \sum_{i=1}^{N_e} e_i(t)\vb{W}_i^{(1)}(\vb{r})$ and
and $\vb{B}(t,\vb{r}) \approx \sum_{i=1}^{N_f} b_i(t)\vb{W}_i^{(2)}(\vb{r})$.
Faraday's law is measured with the 2-form basis function and Ampere's law with the 1-form basis function to yield
\begin{subequations}
\begin{equation}
\partial_t \vb{b} = [\mathcal{L}_{\curl}\circ]\vb{e}
\end{equation}
\begin{equation}
\partial_t \vb{e} = c^2[\mathcal{L}_{\curl}\circ]^T[\star_{\mu^{-1}}]\vb{b} -\varepsilon^{-1}\vb{j}
\end{equation}
\end{subequations}
assuming the domain is freespace with the speed of light $c=\sqrt{\varepsilon_0\mu_0}^{-1}$.
The Hodge operator matrix is defined as
\begin{equation}
[\star_{\mu^{-1}}]_{ij} = \int_\Omega d\Omega \vb{W}_i^{(2)}(\vb{r})\cdot\vb{W}_j^{(2)}(\vb{r}).
\end{equation}
The degrees of freedom for the current are defined as
\begin{equation}
    \vb{j}_i = \sum_{n=1}^{N_p}\frac{q_n}{\Delta_t}\int_{\vb{r}^n}^{\vb{r}^{n+1}}d\vb{p}_n\cdot \vb{W}_i^{(1)}(\vb{r})
\end{equation}
where $\vb{p}_n$ is the particle path for the $n$-th particle.

\subsubsection{DH-MFEM Formulation\label{app:DH}}

In this formulation, $\vb{H}(t,\vb{r}) \approx \sum_{i=1}^{N_e} h_i(t)\vb{W}_i^{(1)}(\vb{r})$ and $\vb{D}(t,\vb{r}) \approx \sum_{i=1}^{N_f} d_i(t)\vb{W}_i^{(2)}(\vb{r})$.
Ampere's law is measured with the 2-form basis function and Faraday's law with the 1-form basis function to yield
\begin{subequations}
\begin{equation}
\partial_t \vb{h} = c^2[\mathcal{L}_{\curl}\circ][\star_{\varepsilon^{-1}}]\vb{d}
\end{equation}
\begin{equation}
\partial_t \vb{d} = [\mathcal{L}_{\curl}\circ]\vb{h} -\vb{j}.
\end{equation}
\end{subequations}
The Hodge operator matrix is defined as
\begin{equation}
[\star_{\varepsilon^{-1}}]_{ij} = \int_\Omega d\Omega \vb{W}_i^{(2)}(\vb{r})\cdot\vb{W}_j^{(2)}(\vb{r}).
\end{equation}
The degrees of freedom for the current are defined as
\begin{equation}
    \vb{j}_i = \sum_{n=1}^{N_p}\frac{q_n}{\Delta_t}\int_{\vb{r}^n}^{\vb{r}^{n+1}}d\vb{p}_n \hat{n}_i\cdot \vb{J}_n(\vb{r})
\end{equation}
where the particle path for the $n$-th particle and is only evaluated on the surface.
\normalem

\nocite{*}
\bibliography{aipsamp}

\end{document}